\documentclass[12pt, a4paper,titlepage]{article}

\usepackage{amsmath}

\usepackage{graphicx}
\usepackage{color}
\usepackage{booktabs}
\usepackage[font={small,it}]{caption}
\usepackage[round,numbers,sort&compress]{natbib} 

\newcommand {\vct}[1] {{\mathbf{#1}}}
\newcommand {\tnsr}[1] {\bar{\mathsf{#1}}}
\title{Force Spectroscopy with Dual-Trap Optical Tweezers: Molecular Stiffness Measurements and Coupled Fluctuations Analysis }

\author{M. Ribezzi--Crivellari\\
          Departament de Fisica Fonamental,\\
	   Universitat de Barcelona  
\and
F. Ritort \\
	 Departament de Fisica Fonamental,\\
	   Universitat de Barcelona,\\
            \& \\
            Ciber-BBN de Bioingeneria, Biomateriales y Nanomedicina }

\begin{document}

\maketitle

\abstract{
Dual trap optical tweezers are often used in high-resolution measurements in single-molecule biophysics. 
Such measurements can be hindered by the presence of extraneous noise sources, the most prominent of which is the coupling of fluctuations
along different spatial directions, which may affect any optical tweezers setup. In this paper we analyze, both from the theoretical and the experimental
points of view, the most common source for these couplings in Dual Trap Optical Tweezers setups: the misalignment of traps and tether. We give criteria to distinguish 
different kinds of misalignment, to estimate their quantitative relevance and to include them in the data analysis. 
The experimental data is obtained in a novel dual trap optical tweezers setup which directly measures forces.
In the case in which misalignment is negligible we provide a method to measure the stiffness of traps and tether
based on variance analysis. This method can be seen as a calibration technique valid beyond the linear trap region.
Our analysis is then employed to measure the persistence length of ds-DNA tethers of three different lengths
spanning two orders of magnitude. The effective persistence length of such tethers is shown to decrease with the contour length, in accordance with previous studies.
}

\emph{Key words; keyword 1; keyword 2; keyword 3; keyword 4}

\clearpage

\section*{Introduction}

Optical Tweezers (OT) have been often employed to measure the elastic properties of polymers tethered between dielectric beads. 
A direct measurement of the tether stiffness is possible through fluctuation analysis \cite{Prl.meiners.2000}.
This kind of measurements are appealing from the experimental point of view because they require the measurement of a single
quantity, either force or extension, so that they do not need a prior determination of the trap stiffness. 
One major source of error in these measurements is the coupling of fluctuations along different
spatial directions, mainly in and out of the focal plane. This is a nontrivial effect which is expected to affect, 
although to different extents, any OT setup. The relevance of such effect is not limited to fluctuation measurements: it can also affect 
the correct measurement of force-extension curves, especially for short tethers.
A clear understanding of the physical basis of such couplings is useful to determine under which conditions they 
lead to systematic errors in the measurements. 
In this paper we study fluctuation coupling in Double-Trap Optical Tweezers (DTOT) setups which are currently used in
high resolution force spectroscopy. The most important coupling sources are misalignmnt effects affecting both traps and tether. 
Through theoretical modeling we establish a criterion to identify the kind of misalignment which causes the coupling and give explicit formulas
to quantify its effect on the basis of the parameters characterizing the experimental setup. These considerations are also
relevant for Single Trap Optical Tweezers (STOT).
Moreover we show that, if coupling effects are negligible, the analysis of the variance of the measured signals, either
force or position, may be used to simultaneously measure the tether and trap stiffnesses, including possible asymmetries between the
two traps. This kind of measurements can be used as a calibration technique, suitable for any DTOT, which works beyond the linear region of the traps. 
When instead couplings are non-negligible we show how to include them in the data analysis.
To illustrate this methodology we carried out measurements in a novel DTOT setup which uses counterpropagating
beams and measures forces directly using linear momentum conservation. This system is especially well suited for 
stiffness measurements as the force measurement calibration is independent of the shape and size of the trapped object. 
This is not true when using other techniques, e.g. back focal plane interferometry, which require a specific calibration of the position measurement for 
each bead. We performed direct measurements of the stiffness of ds-DNA tethers whose contour length spans two decades.

\section*{Materials and Methods}
\begin{figure}
 \centering
\includegraphics[width=4in]{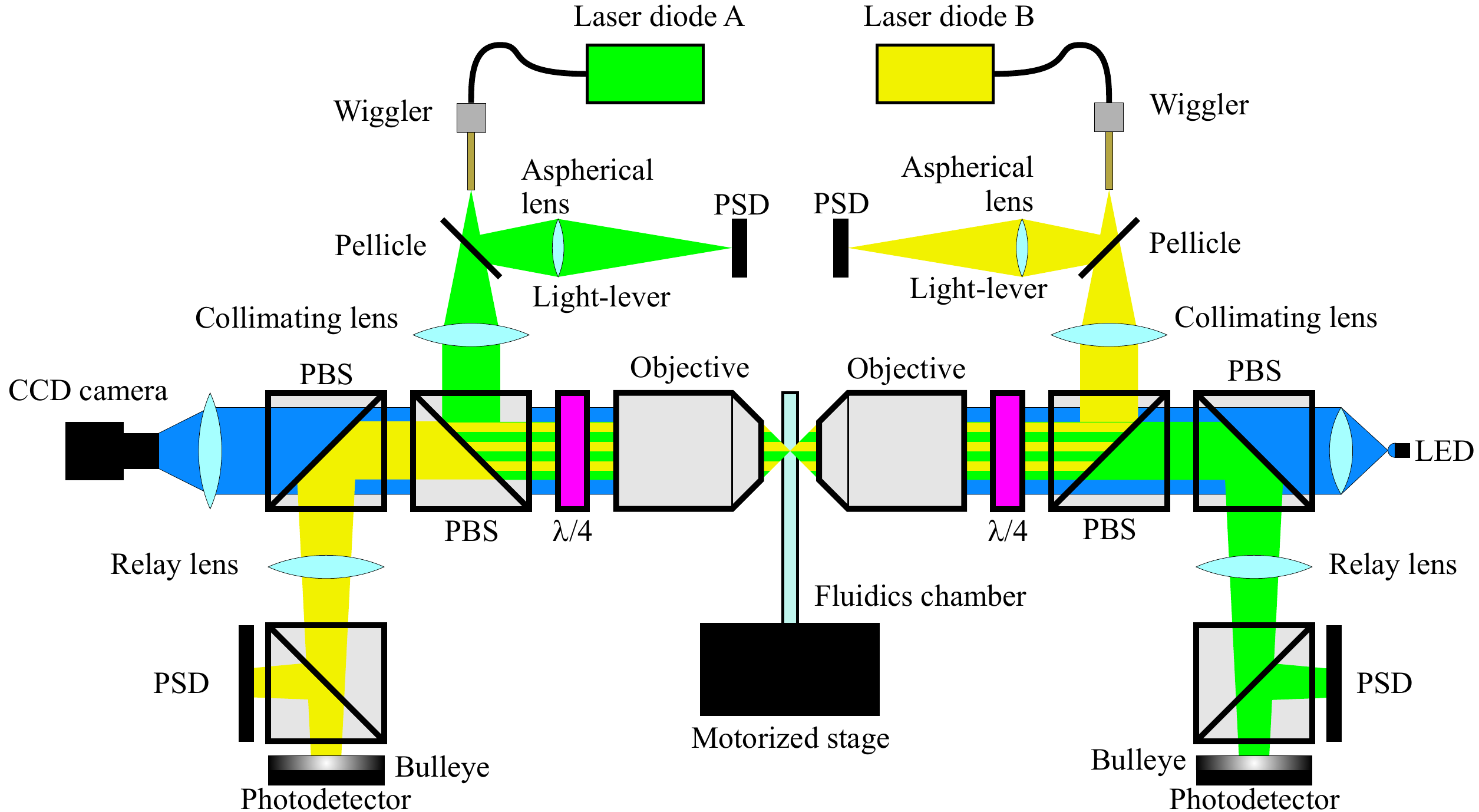}
\caption{\label{Figure:opts}}
\end{figure}

\begin{small}
\subsection*{Optical Tweezers Setup}

The DTOT setup is shown in Figure \ref{Figure:opts} and is very similar to the one designed by Smith et al. \cite{MEnzy.smith.2003} 
and described in \cite{PNAS.huguet.2010}, which operates with a single trap and a pipette.
Force measurement is based on the 
conservation of linear momentum \cite{MEnzy.smith.2003}, making force calibration very robust. Calibration factors are determined by the optical setup 
and the detector response but they are independent of other details of the experimental setup, such as the index of refraction of the trapped object, its size or shape, the refractive 
index of the buffer medium and laser power. Unless the optics or the detector are changed, there is no need for continued calibration \cite{MEnzy.smith.2003}.
Fluctuation measurements were performed using an acquisition board (Agilent Technologies) with a 50 kHz bandwidth, which is higher than corner frequencies 
of the measured signals. In a typical experiment fluctuations were measured for 10 seconds, by increasing the force in 2 pN steps between subsequent measurements.

\subsection*{Molecular synthesis}

For our experiments we used ds-DNA tethers of four different lengths. 
The 24 kb tether was obtained by digesting the phage-$\lambda$ plasmid with the XbAI restriction enzyme. 
The tether was then ligated to a biotin-labeled oligo on one side and with a dig labeled oligo on the other side.
The 3 kb tether was obtained by PCR amplification of a section of the phage-$\lambda$ plasmid using biotin-modified primers. The amplified segment
was then restricted with Xba I and ligated to a dig-modified oligo. 
The 1.2 kb and 58 b tethers were synthesized according to the protocol described in \cite{BiophysJ.forns.2011}.
Experiments were performed in a microfluidic chamber formed by two coverslips interspaced with parafilm. Anti-dig coated beads
were first incubated with the molecule of interest and then introduced one at a time in the microfluidics chamber trough a dispenser tube. Once the
anti-dig coated bead was trapped a streptavidin coated bead was introduced through a second dispenser tube and trapped in the second trap.
The connection was then formed directly inside the microfluidics chamber. 
All experiments on DNA tethers were performed in PBS buffer solution at pH 7.4, NaCl 1M, at $25^\text{o}$C. 
This buffer solution was found to greatly reduce the nonspecific adsorption of DNA on silica. 
We dissolved $1mg/\mu l$ Bovine Serum Albumin in the buffer in order to reduce  non-specific
silica-silica interactions.

\end{small} 

\section*{Results}

\subsection*{Coupled fluctuations in a plane.}\label{coupletwo}

\begin{figure}
\centering
\includegraphics[width=4in]{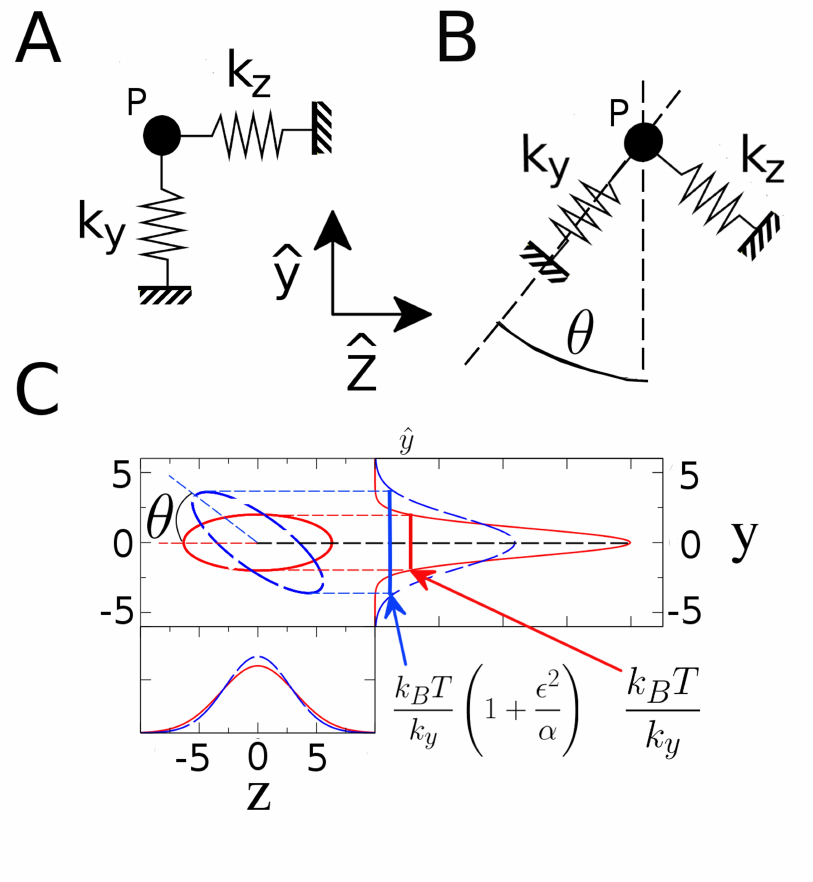} 
\caption{\label{Fig.s}}
\end{figure}

In order to introduce the main subject of this paper it is useful to consider a pedagogical example that shows how a small coupling between two fluctuating degrees of freedom can have a large effect on their variances. 
Let us consider a particle moving on a plane while constrained by two harmonic springs (Fig. \ref{Fig.s} A). The position of the particle is given by a vector $\vct{p}=(y,z)$ 
and the springs are oriented along the coordinate axes ($\hat y,\hat z$) (Fig.\ref{Fig.s}A).
The stiffness matrix $\tnsr{k}$ is given by: 
\begin{equation}
 \tnsr{k}=\left(\begin{array}{c c}
             k_y & 0 \\
             0 & k_z
            \end{array}\right)=k_y \left(\begin{array}{c c}
             1 & 0 \\
             0 & \alpha
            \end{array}\right),\, \alpha=k_z/k_y,
\end{equation}
where $k_y$,$k_z$ are the stiffnesses along $y$,$z$ respectively.
The equilibrium Boltzmann distribution for $\vct{p}$ can be written in compact form as:
\begin{eqnarray}\label{pocci}
\rho(\vct{p})&=&\frac{1}{Z}\exp\left(-\frac{\vct{p}\cdot \tnsr{k}\vct{p}}{2k_BT}\right)\\
Z&=&\int dzdy \exp\left(-\frac{\vct{p}\cdot \tnsr{k}\vct{p}}{2k_BT}\right),
\end{eqnarray}
where $k_B$ is the Boltzmann constant, $T$ the absolute temperature and $Z$ the partition function. The variance of $\vct{p}$ is thus:
\begin{equation}
\langle\vct{p}^2\rangle =k_BT  \tnsr{k}^{-1}\simeq\frac{k_BT}{k_y}\left(\begin{array}{c c} 1 & 0\\
                                                                  0 & \frac 1 \alpha
                                                                  \end{array}\right).
\end{equation}
In the case $k_y\gg k_z$ ($\alpha\ll 1$) the variance of spatial fluctuations along $z$ is much bigger than the variance along $y$ and the level curves 
of the probability distribution in Eq. \eqref{pocci} are highly eccentric ellipses (Fig. \ref{Fig.s} C, red solid curves). 

If the springs are misaligned by an angle $\theta$ with respect to the reference frame (Fig. \ref{Fig.s} B), the stiffness tensor changes as:
\begin{equation}\label{sannio}
\begin{split}
&\tnsr{k}'=\tnsr{R}(\theta) \tnsr{k} \tnsr{R}^T(\theta)=\\
=&\left(\begin{array}{c c} k_y \cos^2\theta+k_z \sin^2\theta & (k_y-k_z)\cos\theta\sin\theta \\  (k_y-k_z)\cos\theta\sin\theta & 
k_z \cos^2\theta+k_y \sin^2\theta
                                                                  \end{array}\right).
\end{split}
\end{equation}
where $\tnsr{k}'$ is the new stiffness and $\tnsr{R}(\theta)$ is a rotation of angle $\theta$.
In this situation, since $k_y\neq k_z$, the off--diagonal terms in Eq. \eqref{sannio} couple the motion of the particle along $y$ and $z$. We 
 define $\epsilon=\sin\theta$ as the coupling parameter. 
To lowest order in $\epsilon$ we get:
\begin{equation}\label{reppa}
\begin{split}
 \langle\vct{p}^2\rangle &=k_BT  \tnsr{k'}^{-1}\simeq\\
&\simeq \frac{k_BT}{k_y}\left(\begin{array}{c c} 1+\left(\frac 1 \alpha-1 \right)\epsilon^2 & \epsilon\left(1-\frac 1 \alpha\right) \\
                                                                  \epsilon\left(1 -\frac 1 \alpha\right) & \frac 1 \alpha \left(1-\epsilon^2\right)+\epsilon^2 
                                                                  \end{array}\right).
\end{split}
\end{equation}
For a small coupling $\epsilon$ the variance of $y$ is increased to:
\begin{equation}
\begin{split}
 \langle y^2\rangle&= \frac{k_BT}{k_y}\left(1+\left(\frac{1}{\alpha}-1\right){\epsilon^2}\right)\simeq\\
&\simeq \frac{k_BT}{k_y}\left(1+\frac{\epsilon^2}{\alpha}\right),\qquad (\alpha \ll 1).
\end{split}
\end{equation}

This shows that the effect of a small coupling $\epsilon$ (small misalignment of the springs) on the variance of $y$
can be large if $\frac{\epsilon^2}{\alpha}\sim 1$. In other words, the small $\alpha$ value acts as a lever arm, amplifying the effect of
a small rotation (Fig. \ref{Fig.s} C). 
Only when
\begin{equation}\label{condition}
\epsilon^2\ll \alpha\qquad,
\end{equation}
the coupling of fluctuations can be ignored.
So far we gave the general picture of the effect of a coupling between two fluctuating degrees of freedom. 
In the next section we will show how this simple model
can be extended to study fluctuations in a DTOT. Generally speaking, when a molecule is pulled at high forces the rigidity of the 
tether is much higher along the pulling direction ($y$) 
than in the perpendicular direction ($z$) and Eq.\eqref{condition} can be violated even if $\epsilon$ is very small ($\simeq 0.1$).
At low forces and for long tethers Eq.\eqref{condition} is instead fulfilled and the coupling can be ignored. 
A quantitative treatment of these effects and the corresponding data analysis are described next.

\subsection*{Coupled fluctuations in a DTOT setup}

\begin{figure}
\centering\includegraphics[width=3.25in]{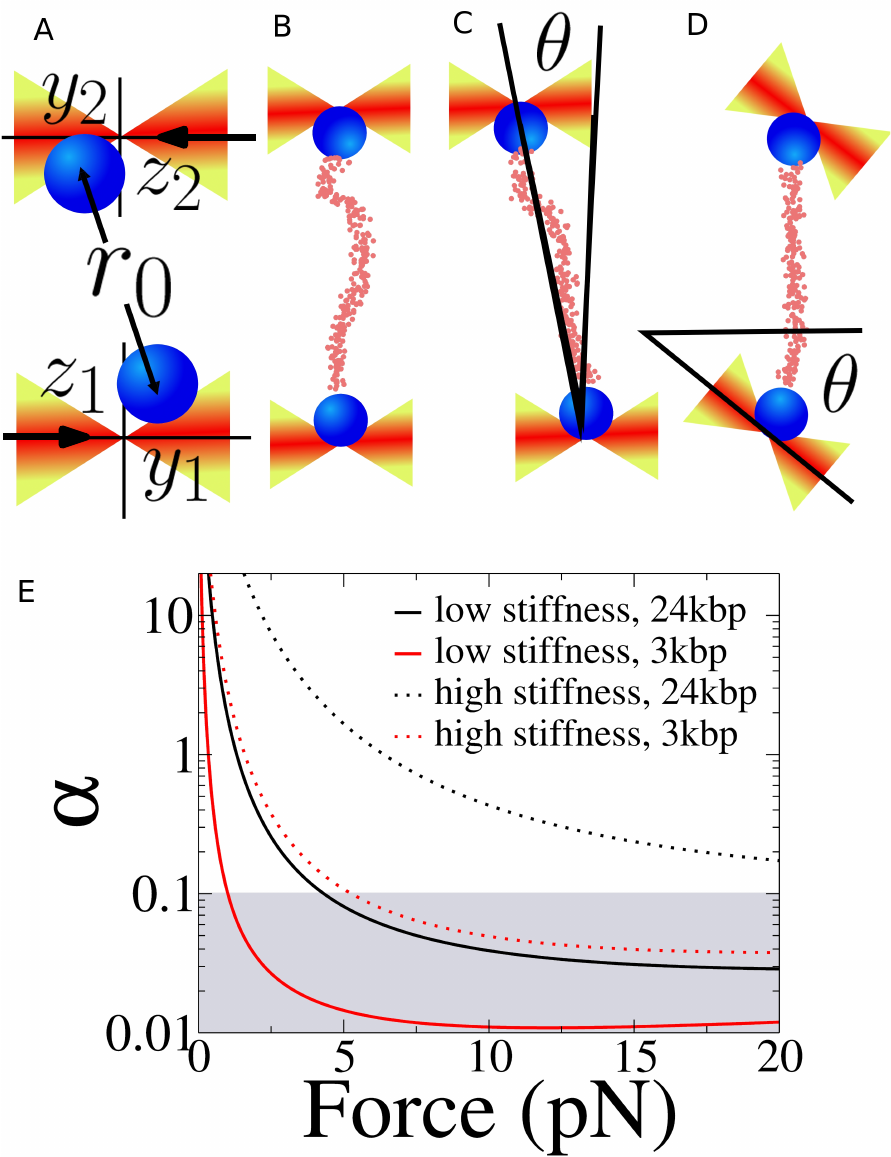} 
\caption{\label{Fig:setup}}
\end{figure}

Correlation functions of fluctuation measurements in a DTOT do often show a double exponential nature. This 
is usually interpreted in terms of a coupling of the fluctuations in the optical plane and along the optical axis 
\cite{Prl.meiners.1999,CSP.bustamante.2009}. Couplings affect the variance of the
measured force (or position) signal and the spurious contribution must be removed from the measurements. Yet, coupling effects reflect 
some defect in the design of the experimental setup and their study can provide valuable diagnostic tools for the fine
tuning of a DTOT setup, crucial for high resolution measurements. Most prominent coupling sources are misalignment effects, due to at least two causes. 
A first cause is tether misalignment, i.e.  
a configuration in which the centers of the traps lie in different planes (Fig. \ref{Fig:setup}C). A second case is trap misalignment, when the principal 
axes of the traps are tilted with respect to the direction along which the tether is stretched (Fig. \ref{Fig:setup}D). 
Remarkably these two scenarios lead to different coupling structures making possible the identification of misalignment effects.
To discuss coupling effects it is necessary to consider a 4-dimensional configuration space, describing the position of the beads  both along the optical 
axis and the pulling direction.
The laser beams (black arrows in Fig. \ref{Fig:setup}A) define the optical axis, which we identify with the $\hat{z}$ direction 
in our reference frame. 
In the optical plane (perpendicular to the optical axis), we shall use the coordinate $\hat{y}$ to denote the direction along which the tether is oriented. 
The coordinate $\hat{x}$, perpendicular to both $\hat{y}$ and $\hat{z}$, will play no role in our analysis. 
The effect of both traps and tether on the dynamics of the beads can be modeled by a potential energy function,
which depends on the positions of the beads in the $y-z$ plane. 
When considering equilibrium fluctuations at constant trap--to--trap distance, a linear approximation around the equilibrium positions can be used:
\begin{equation}
U(y_1,y_2,z_1,z_2)=\frac{1}{2} \vct{p}^T\tnsr{K'}\vct{p}, 
\end{equation}
with $\vct{p}=(y_1,y_2,z_1,z_2)$, and $\tnsr{K'}$ the stiffness tensor (tensors are primed when they are represented in the $y-z$ coordinate system).
In the ideal case (Fig. \ref{Fig:setup}B), the stiffness tensor, which is now $4\times4$, reads:
\begin{equation}\label{r}
 \tnsr{K'}=\bordermatrix{~ & y_1 & y_2 & z_1 & z_2 \cr
                  y_1 & k_y+k_m & -k_m & 0 & 0 \cr
                  y_2 & -k_m & k_y+k_m & 0 &0 \cr
                  z_1 & 0 & 0 & k_z+\frac{f}{r_0} &-\frac{f}{r_0} \cr
                  z_2 & 0 & 0 & -\frac{f}{r_0} & k_z+\frac{f}{r_0} \cr
                }.
\end{equation}

The stiffness tensor is two-block diagonal, with the first block describing the effect of traps and tether along the $y$ axis, 
and the second block describing the effect of traps
and tether in the $\hat z$ direction, sharing the same structure. Here $k_y$ is the stiffness of the trap in the $\hat y $ direction (the two traps
are assumed identical for simplicity), $k_z$ is the trap stiffness in the $\hat z$ direction, $k_m$ is the stiffness of the tether connecting the
two beads, $f$ is the mean force and $r_0$ the mean distance between the centers
of the beads.
In this case the stiffness tensor can be diagonalized switching to the coordinate system defined by:
\begin{eqnarray}
y_+=\frac{y_1+y_2}{\sqrt{2}},\qquad y_-=\frac{y_1-y_2}{\sqrt{2}}\\
z_+=\frac{z_1+z_2}{\sqrt{2}},\qquad z_-=\frac{z_1-z_2}{\sqrt{2}},
\end{eqnarray}
where $y_+,z_+$ represent the center of mass position and $y_-,z_-$ represent the differential coordinate.
In this second coordinate system, the stiffness tensor reads:
\begin{equation}\label{casoideale}
 \tnsr{K}= \bordermatrix{~ & y_+ & y_- & z_+ & z_- \cr
                  y_+ & k_y & 0 & 0 & 0 \cr
                  y_- & 0 & k_y+2k_m & 0 &0 \cr
                  z_+ & 0 & 0 & k_z &0 \cr
                  z_- & 0 & 0 & 0 & k_z+2\frac{f}{r_0} \cr
                },
\end{equation}
which shows that the four different fluctuation modes are uncoupled (we dropped the prime because we switched to a different reference frame).
The tensor $\tnsr{K}$ can be written as the sum of two contributions:
\begin{equation}
\tnsr{K}=\tnsr{K}_T+2\tnsr{K}_m, 
\end{equation}
with:
\begin{equation}
\tnsr{K}_T=\left( \begin{array}{ c c c c} k_y & 0 & 0 & 0\\
                    0 & k_y & 0 & 0 \\
                    0 & 0 & k_z & 0\\
                    0 & 0 & 0 & k_z
                   \end{array} \right),
\end{equation}
which accounts for the trap contribution to the stiffness, and:
\begin{equation}\label{cappaemme}
\tnsr{K}_m=\left( \begin{array}{ c c c c} 0 & 0 & 0 & 0\\
                    0 & k_m & 0 & 0 \\
                    0 & 0 & 0 & 0\\
                    0 & 0 & 0 & \frac{f}{r_0}
                   \end{array} \right),
\end{equation}
which accounts for the tether contribution to the stiffness.
\subsubsection*{Tether misalignment effects}

In the presence of tether misalignment, i.e. if the two traps are focused at different depths along the optical axis (Fig. \ref{Fig:setup}C), the tether forms an angle $\theta$
with respect to the $y$ axis. This can be incorporated in the stiffness tensor through a rotation of $\tnsr{K}_m$, Eq. \eqref{cappaemme}, of the same angle:
\begin{equation}
\tnsr{K}_m(\epsilon)=\left( \begin{array}{ c c c c} 0 & 0 & 0 & 0\\
                    0 & u(\epsilon) & 0 & \epsilon w(\epsilon) \\
                    0 & 0 & 0 & 0\\
                    0 & \epsilon w(\epsilon) & 0 & v(\epsilon)\end{array} \right),
\end{equation}
with  
\begin{eqnarray}
u(\epsilon)&=&k_m (1-\epsilon^2)+\frac{f}{r_0} \epsilon^{2}\\
v(\epsilon)&=&\frac{f}{r_0} (1-\epsilon^2)+k_m \epsilon^2\\
w(\epsilon)&=&\left(k_m-\frac{f}{r_0}\right) \sqrt{1-\epsilon^2},
\end{eqnarray}
and $\epsilon=\sin(\theta)$.

The total stiffness tensor is now:
\begin{equation}\label{fort}
 \tnsr{K}(\epsilon)= \bordermatrix{ ~    & y_+   & y_-                  &  z_+  &   z_-                  \cr
                          y_+   & k_y   & 0                    &  0    &   0                    \cr
                          y_-   &  0    & k_y+2u(\epsilon)     &  0    &   \epsilon w(\epsilon) \cr
                          z_+   &  0    & 0                    &  k_z  &   0                    \cr                          
                          z_-    &  0    & \epsilon w(\epsilon) &  0    &   k_z+2 v(\epsilon)    \cr                                
                          },
\end{equation}
which must be compared to Eq. \eqref{casoideale}.
Looking at the non diagonal terms of this tensor it is evident that the coupling will only affect the two differential modes, $y_-,z_-$ 
leaving the motion of the center of mass, $y_+,z_+$, unchanged.

\subsubsection*{Trap misalignment effect}
In the second configuration the trap stiffness tensor is rotated (Fig. \ref{Fig:setup}D), and the stiffness tensor becomes:
\begin{equation}
 \tnsr{K}(\epsilon)= \bordermatrix{~ & y_+ & y_- & z_+ & z_- \cr
                  y_+ & p(\epsilon) & 0 &  \epsilon q(\epsilon) & 0 \cr
                  y_- & 0 & p(\epsilon)+2k_m & 0 & \epsilon q(\epsilon) \cr
                  z_+ & \epsilon q(\epsilon) & 0 & m(\epsilon) & 0 \cr
                 z_- & 0 & \epsilon q(\epsilon) & 0 & m(\epsilon)+2\frac{f}{r_0} \cr
                }.
                \end{equation}
with 
\begin{eqnarray}
p(\epsilon)&=&k_y (1-\epsilon^2)+k_z \epsilon^{2}\\
m(\epsilon)&=&k_z (1-\epsilon^2)+k_y \epsilon^2\\
q(\epsilon)&=&\left(k_y-k_z\right) \sqrt{1-\epsilon^2},
\end{eqnarray}
and $\epsilon=\sin(\theta)$.
In contrast to the previous case, the non-diagonal terms show that the coupling will affect both the differential and center of mass coordinates.
This important difference provides a simple tool to distinguish between tether and trap misalignments.

\subsection*{On the relevance of coupling effects}
In the case of tether misalignment the off diagonal terms in Eq. \eqref{fort} couple the fluctuations along the $y_-$ and $z_-$ 
directions. A reduced description, addressing these two coordinates is possible using a subtensor of $\tnsr{K}(\epsilon)$, \eqref{fort}:
\begin{equation}\label{fortx}
 \tnsr{K}_-(\epsilon)= \bordermatrix{ ~    & y_-     &   z_-           \cr
                          y_-   & k_y+2u(\epsilon)      &   \epsilon w(\epsilon) \cr                          
                          z_-    & \epsilon w(\epsilon) &   k_z+2 v(\epsilon)    \cr                                
                          }.
\end{equation}
The variance of $y_-$ is given by:
\begin{equation}\label{cozza}
 \langle y_-^2\rangle=k_BT\left(\tnsr{K}'_-(\epsilon)^{-1}\right)_{y_-y_-},
\end{equation}

which, to leading order in $\epsilon$, is:
\begin{equation}\label{gg}
\frac{\langle y_-^2\rangle}{\langle y_-^2\rangle_{\epsilon=0}}=1+\frac{\epsilon^2}{\alpha} 
\end{equation}
with
\begin{equation}
\alpha=\frac{(k_y+2k_m)(k_z+2f/r_0)}{(2 k_m+k_z)(2 k_m-2f/r_0)}.
\end{equation}


The value of $\alpha$ can be computed if the stiffnesses are known. In Fig. \ref{Fig:setup} E we show the behavior of $\alpha$ as a function of the
force for two different ds-DNA tethers (3kbp and 24 kbp) and for two different trap stiffnesses. The continuous curves show the value of $\alpha$ computed 
in a low stiffness  setup ($k_y=0.02$ pN/nm, $k_z=0.001$ pN/nm, the condition in which our DTOT operates) 
while the dotted lines show the value of $\alpha$ for high trap stiffness 
($k_y=0.2$ pN/nm, $k_z\simeq0.01$ pN/nm, as reported for the DTOT in \citep{PNAS.gebhardt.2010,Nature.comstock.2011}).
Even in the case of high trap stiffness the attained value of $\alpha$ is small ($\simeq 10^{-1}$) at high forces for the shorter tether. In Fig. \ref{Fig:setup} E
the shaded area shows the
values of $\alpha$ for which a misalignment $\epsilon\simeq 0.1$ causes a 10\% error in the measurements of $y_-$ fluctuations Eq. \eqref{gg}. In our DTOT setup such coupling in negligible when 
using long molecules (8 $\mu$m) at low and moderate forces (up to $\simeq$10 pN).

In the case of trap misalignment both the variances of $y_-$ and $y_+$ are affected and similar formulas hold:
\begin{equation}
 \frac{\langle y_-^2\rangle}{\langle y_-^2\rangle_{\epsilon=0}}=1+\frac{\epsilon^2}{\gamma}\qquad\frac{\langle y_+^2\rangle}{\langle y_+^2\rangle_{\epsilon=0}}=1+\frac{\epsilon^2}{\delta}
\end{equation}

\begin{equation}\label{rr}
\gamma=\frac{(k_y+2k_m)(k_z+2f/r_0)}{(k_y-k_z)(k_y+2f/r_0)}
\end{equation}

\begin{equation}\label{tt}
\delta=\frac{k_z}{k_y-k_z}.
\end{equation}

Equations \eqref{gg},\eqref{rr},\eqref{tt} estimate the error due to misalignment as a function of the stiffnesses contributing to the
experimental setup (Fig. \ref{Fig:setup}E). As such they are useful to understand in which force regimes misalignment is going to be important.

\subsection*{Neglecting coupling effects: variance analysis}\label{stifres}

\begin{figure}
\centering\includegraphics[width=3.25in]{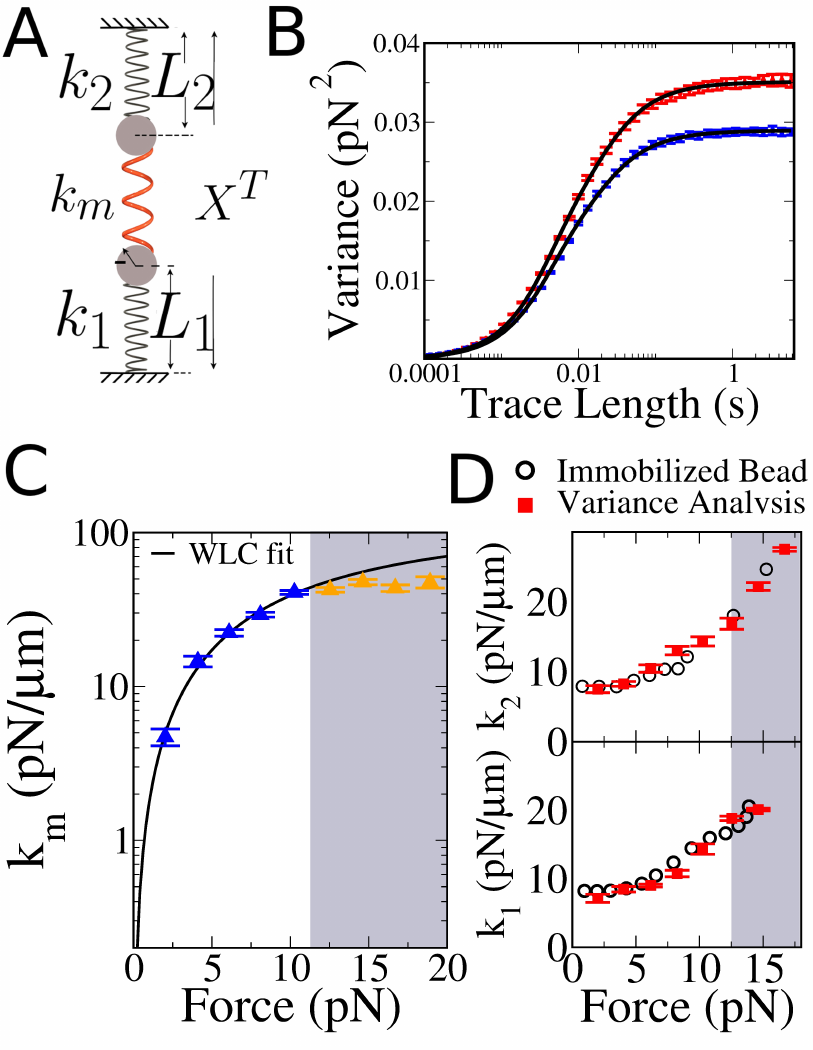} 
\caption{\label{Fig:DNAstiff}}
\end{figure}

In experimental setups where force is 
directly measured, the trap stiffness is used to measure the extension of the fiber. Conversely, when the bead displacement is measured 
(e.g. by video microscopy or by back focal plane interferometry \citep{RevSciInst.neuman.2004}), the stiffness
is used to measure force. In both cases the trap stiffness depends on the details of the experimental setup i.e. laser power, bead size and shape or buffer medium 
(via its refraction index). Optical traps are assumed to be linear close to the trap center, so that a stiffness measurement at zero force can be used
to characterize the trap shape in this region. This kind of calibration has been shown to achieve 1\% accuracy \cite{RevSciInst.tolic.2006}. Nevertheless when one needs 
to do measurements at high forces, nonlinear effects may become relevant, especially at low trap power and the full force field of the trap should be measured \cite{OptLett.jahnel.2011}. 
When the coupling is negligible, motions along the $\hat y$ and $\hat z$ directions are uncoupled and the stiffnesses of traps and tethers can be obtained
from a straightforward analysis of the experimental force (or position) variances.
We will model traps and thether by the dumbbell shown in Fig. \ref{Fig:DNAstiff}. It consists of three serially connected 
harmonic springs. In this setting we allow for a different stiffness in the two traps: $k_1$, $k_2$ (Fig. \ref{Fig:DNAstiff} A).
Together with the tether stiffness $k_m$ we get the stiffness tensor:
\begin{equation}
 \tnsr{K}'=\left( \begin{array}{ c c } k_1+k_m & -k_m \\
                    -k_m & k_2+k_m 
                   \end{array} \right),
\end{equation}
which is just the first block of Eq. \eqref{r}, and the potential energy of the system is written as:
\begin{equation}
 U(y_1,y_2)=\frac{1}{2}(y_1,y_2)\cdot \tnsr{K}'(y_1,y_2)
\end{equation}
where $(y_1,y_2)$ have the same meaning as in the previous sections.
The equilibrium distribution for $(y_1,y_2)$ is related to $\tnsr{K}'$ by:
\begin{equation}
P(y_1,y_2)=\frac{1}{Z}\exp\left(- \frac{U(y_1,y_2)}{k_BT}\right).
\end{equation}
The variances and covariance of $(y_1,y_2)$ are linked to the inverse of the stiffness tensor.
In terms of the experimentally measured variances and covariances: 
\begin{eqnarray}
k_1 &=&\kappa\frac{\langle y_2^2\rangle-\langle y_1 y_2\rangle}{k_BT},\label{p1}\\
k_2 &=&\kappa\frac{\langle y_1^2\rangle-\langle y_1 y_2\rangle}{k_BT},\label{p2}\\
k_m &=&\kappa\frac{\langle y_1 y_2\rangle}{k_BT}\label{p3},
\end{eqnarray}
with $\kappa^{-1}=\frac{\langle y_1^2\rangle \langle y_2^2\rangle- \langle y_1 y_2\rangle ^2}{(k_BT)^2}$. 
    
In experimental set-ups which directly measure forces different formulas apply: the experimental values ($\sigma^{2}_{11},\sigma^{2}_{22},\sigma^{2}_{12}$, $\sigma^2_{ij}=\langle f_if_j\rangle-\langle f_i\rangle \langle f_j\rangle,\,\, i=1,2$) and the model parameters ($k_1,k_2,k_m$) are related by:
%
\begin{eqnarray}
 k_1&=&\frac{\sigma_{11}^2+\sigma_{12}^2}{k_B T}\label{ia}\\
k_2&=&\frac{\sigma_{22}^2+\sigma_{12}^2}{k_BT}\label{ib}\\
k_m&=&\frac{1}{k_BT}\frac{\sigma_{12}^2\left( \sigma_{11}^2+\sigma_{12}^2\right)\left(  \sigma_{22}^2+\sigma_{12}^2\right)} { \sigma_{11}^2\sigma_{22}^2-\sigma_{12}^4}\label{ic}.
\end{eqnarray}

The method we have introduced in this section is similar to the one used by Meiners and Quake \citep{Prl.meiners.2000}, the difference being that here we use equal--time 
force covariance whereas \citep{Prl.meiners.2000} uses time--dependent correlation functions. Both methods can be
affected by the presence of extraneous noise sources. The effect of  low frequency noises (such as line noise, drift or air flows) can be
minimized by computing the variance of the force (or distance) signal on traces which are much longer than the corner frequency of the dumbbell, but short
enough so that the low frequency noise sources never dominate the power spectrum (Fig. \ref{Fig:DNAstiff}B). 
The method just presented can be used in any DTOT setup to calibrate the two traps (by measuring $k_1,k_2$) at different forces. Equations \eqref{p1},\eqref{p2} and
\eqref{p3} are useful in experimental set-ups which directly measure bead positions, while Equations \eqref{ia},\eqref{ib},\eqref{ic} refer to set-ups
which directly measure forces. 
Fig. \ref{Fig:DNAstiff}C shows the results of direct stiffness measurements, based on equations \eqref{ia},\eqref{ib},\eqref{ic} on a 24kbp ds-DNA molecule,
in the shaded region misalignment effects gain importance and the measured stiffness departs from the WLC fit obtained at low forces.
Note that the effect of tether misalignment on trap stiffness measurements is negligible (Fig. \ref{Fig:DNAstiff}D). This happens 
because the measurement of trap stiffness, in the absence of large asymmetries between the traps, does only depend on the variance of the center-of-mass coordinate.
In our setup a force dependent trap stiffness calibration appears to be crucial if we want to measure the force vs molecular 
extension curve as the trap response is strongly non-linear (Fig. \ref{Fig:DNAstiff}). This is also the case for many other DTOT and STOT set-ups, especially at high enoughforces.

\subsection*{Measurement of the molecular stiffness}

\begin{figure}
 \centering
\includegraphics[width=4in]{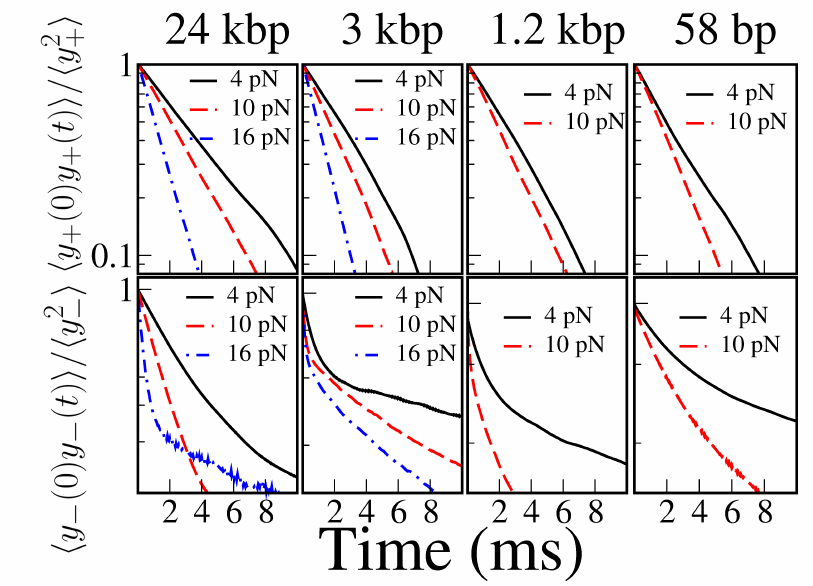}
\caption{\label{Fig:2}}
\end{figure}

The method discussed in the previous section is only useful if coupling effects are negligible, i.e. for long molecules ($10^4$ bp) at small forces ($<10$ pN).
To extend the applicability of direct stiffness measurements to shorter molecules or wider force ranges coupling effects must be taken into account.
Since removing couplings while performing experiments is generally unpractical, it may be convenient to include them data analysis.
To illustrate how this is done we will use fluctuation  measured on ds--DNA tethers of four different lengths: two sections
of the phage--$\lambda$ genome of 24 kb and 3 kb respectively, and two tethers (1.2 kb and 58 b) which are used as handles in single molecule experiments
\cite{BiophysJ.forns.2011}.
The differential, $y_-$ and center-of-mass coordinate $y_+$ correlation functions, measured in our counter propagating DTOT setup are shown in Fig. \ref{Fig:2}.

\begin{figure}
\centering
 \includegraphics[width=3.1in]{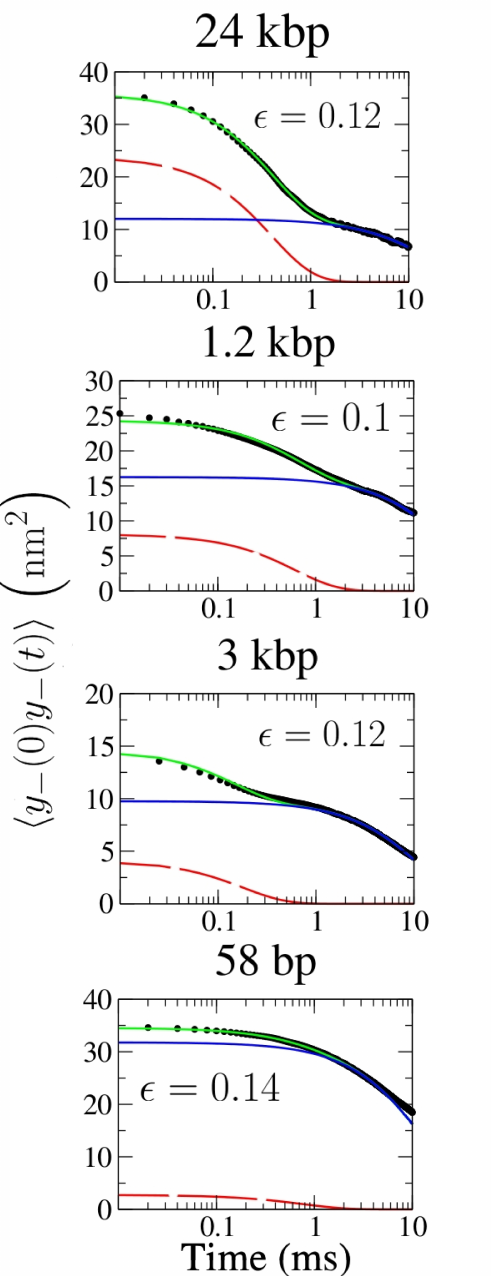}
\caption{\label{Fig:3}}
\end{figure}

\begin{figure}
\centering
\includegraphics[width=3.25in]{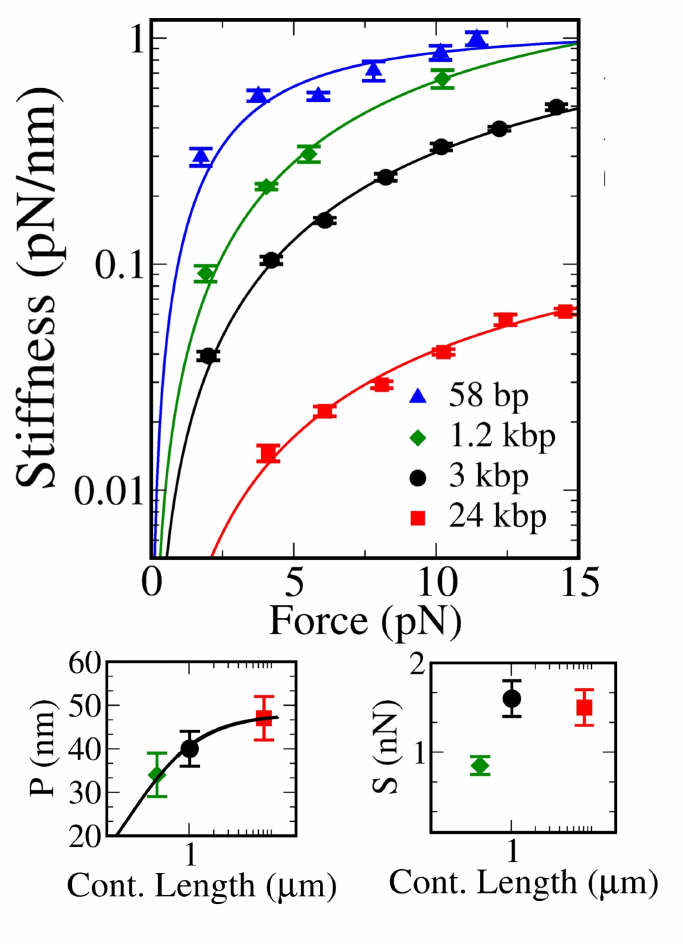} 
\caption{\label{pot}}
\end{figure}

While the correlation function for the center of mass
shows a single exponential behavior (upper panels), the differential coordinate displays a double exponential behavior which derives from the
coupling of fluctuations (lower panels). This is the expected phenomenology for tether misalignment, which is the dominant effect in our setup.
On the contrary, in the case of trap misalignment, both the differential and center of mass coordinates would be affected.
According to Eq \eqref{cozza},\eqref{gg} and neglecting the trap stiffness with respect to the molecular stiffness ($k_m\gg k_y,k_z;f/r_0\gg k_z$), we have that,
in presence of tether misalignment, the variance of the differential coordinate $y_-$ changes to:
\begin{equation}\label{rmod}
\langle y_-^2\rangle \simeq\frac{k_BT}{2k_m}\left(1+\left(\frac{r_0 k_m}{f}-1\right)\epsilon^2\right).
\end{equation}
As the value of $\epsilon$ is not known the derivation of the molecular stiffness cannot be performed on the basis of variance analysis. 
Fortunately, as already noted in \cite{Prl.meiners.1999}, the decay rate of fluctuations in the optical plane $\omega_+$ is much
bigger than that of fluctuations along the optical axis $\omega_-$. The correlation function for $y_-$ (derived in the Supplementary
Material) reads:
\begin{equation}
\langle y_-(0)y_-(t)\rangle=\left(1-\epsilon^{2}\right)\frac{e^{-2\omega_+ t}}{2 k_m}+\epsilon^2 \frac{r_0e^{-2\omega_- t}  }{2f},
\end{equation}
and the molecular stiffness can still be recovered by selecting the amplitude of the fast decaying component of the correlation function (Fig. \ref{Fig:3}).   
Using this data analysis technique we measured the force dependent stiffness of the different tethers (Fig. \ref{pot}). 

The nonlinear elasticity of ds-DNA is usually modeled with the Worm Like Chain (WLC) model. 
When a contribution for enthalpic stretching  and overwinding \citep{Nature.gore.2006} is added, the so--called Extensible WLC (EWLC) \citep{Macromol.marko.1995} model is obtained.
The Marko-Siggia approximation formula for the force dependent stiffness of a EWLC is: 
\begin{equation}\label{EWLC}
 k_m(f)=\frac{1}{\ell_0}\left(\frac{1}{4}\sqrt{\frac{k_bT}{P}}\left(\frac{1}{f}\right)^{3/2}+\frac{1}{S} \right)^{-1},
\end{equation}
where $P$ is the persistence length of the polymer, $\ell_0$ its contour length and $S$ is the stretch modulus. 
This approximation is valid for molecules whose countour length is larger than the persistence length, ($l_0 > P$) \citep{Macromol.marko.1995}.
The EWLC model was fitted to the data letting the persistence length $P$ and the stretch modulus $S$ vary.
Fit results are shown in Table \ref{tab.1}. 
In all cases, the results are compatible with the existing literature, although they were derived in a different force range. In particular the 3kbp molecule shows a decrease in persistence length which is compatible with 
theoretical predictions based on finite-size effects as recently shown in \citep{BiophysJ.seol.2007}.
Data for the 1.2 kbp and 58 bp tethers are instead consistent with measurements performed in a STOT \citep{BiophysJ.forns.2011}.
In principle, 58 bp data should not be fitted with the Marko-Siggia approximation ($l_0<P$) and a semiflexible rod model is required. 
Nevertheless we decided to include the results for such short tether in Figure \ref{pot} to stress the agreement between our dual trap measurements and those reported in 
Ref. \citep{BiophysJ.forns.2011} for a STOT.

Several recent measurements on ds-DNA suggest a strong reduction of the persistence length as the
contour length of the molecule decreases. This could be the result of scale dependent DNA elasticity, as already found in microtubules \citep{PNAS.pampaloni.2006}, or a finite-size effect due to the boundary conditions
\cite{BiophysJ.seol.2007}.
Here we carry out persistence length measurements in the highly stretched regime ($2-15$ pN), i.e. when the molecule is almost fully stretched and 
the enthalpic contribution is important \cite{Macromol.marko.1995,Epl.marko.1997}. In this regime and in contrast to low force measurements \cite{BiophysJ.seol.2007,Pre.chen.2009,BiophysJ.chen.2009}, excluded volume effects between the trapped beads are negligible. 
Moreover the effect of fluctuating boundary conditions on the force-extension curve should become less and less relevant as $fR_b$ grows,
with $f$ the mean tension along the tether and $R_b$ the bead radius ($\simeq$2 $\mu$m in our experiments). 
Seol {\it et al.} propose a phenomenological scaling equation:
\begin{equation}\label{pheno}
P(\ell_0)=\frac{P_\infty}{1+a\frac{P_\infty}{\ell_0}}.
\end{equation}
We fitted such empirical formula to our persistence length measurements, leaving out the shortest thether whose persistence length cannot be correctly measured using
the Marko-Siggia appriximation.  The parameters obtained from the fit are $P_\infty=49\pm2$ nm, in accordance to what measured in 
\cite{BiophysJ.seol.2007} and $a=4\pm1$ which is bigger than the value $(2.78)$  measured in \cite{BiophysJ.seol.2007} and 
later confirmed in \cite{Pre.chen.2009}. It must be stressed that our experiments are performed in a dumbbell configuration, while both \cite{BiophysJ.seol.2007} and
\cite{Pre.chen.2009} have one of the ends of the molecule attached to a surface, and the boundary conditions imposed on the molecule are different in the two cases.
The finite size WLC theory \cite{BiophysJ.seol.2007} does indeed predict a faster decrease of the effective presistence length with the contour length in the 
dumbbell configuration.

\section*{Conclusions}

Misalignment effects may affect any optical tweezers setup and the effect of a small misalignment can be enhanced by the large difference in stiffness
between fluctuations along the pulling direction and along the optical axis. Analyzing these effects we have provided tools to estimate their relevance,
we have shown that trap and tether misalignments lead to different coupling structures and explained how to include couplings in the data analysis. 
When misalignment is negligible, it is possible to measure the stiffnesses of traps and tether from a straightforward analysis of the measured variance
of force (or position) signals. This technique may be used as a calibration technique, valid beyond the linear trap region, in any DTOT. Otherwise it
is possible to include misalignment effects in data analysis.
Such techniques have been used to measure the stiffness of ds-DNA tethers of three different lengths, from 24 kbp to 1200 bp.
The stiffness was interpreted in the framework of EWLC model and we confirmed the decrease in apparent persistence length previously reported in \cite{BiophysJ.seol.2007,Pre.chen.2009},
although our measurements were performed in a wider force range and with a different set-up.


\section*{Financial Support}

FR is supported by  MICINN FIS2007-3454, HFSP Grant No. RGP55-2008, and ICREA Academia grants. MR is supported by HFSP Grant No. RGP55-2008.

\begin{center}
\begin{table}
\centering
\begin{tabular}{|c | c  c |}
\toprule \hline
Molecule &  \multicolumn{2}{c}{Fit Results} \\
 & P (nm) & S (pN) \\
24 kbp & $48\pm 5$ & $1400 \pm 300$ \\
3 kbp & $ 39 \pm 4$ & $ 1800 \pm 400$\\
1.2 kbp & $34\pm 5$ & $850 \pm 100$\\
\hline
58 bp & $1.4 \pm 0.8$ & $20\pm 2$ 
\end{tabular}
\caption{\label{tab.1}{\bf Persistence length and stretch modulus for different ds-DNA tethers.} Errors are standard deviations over different molecules. 
In all cases the measurements were obtained on at least three different molecules. The Marko-Siggia approximation is not valid in the case of the shortest tether
and Eq. \eqref{EWLC} cannot consistently be used to estimate the persistence length of such tether. Nevertheless we included this result to compare it with that
obtained in Ref. \citep{BiophysJ.forns.2011} using a STOT.
The data for the persistence length of two shortest molecules agree with those obtained using
a STOT \citep{BiophysJ.forns.2011}, which are $P=31\pm3$ nm,  for the 1.2kb molecule and $P=1.6\pm0.3$ nm for the 58 bp molecule}
\end{table}
\end{center}

\clearpage
\section*{Figure Captions}

\subsection*{Figure 1}
{\bf Experimental Setup.} The scheme of the optical setup, with the optical paths of the lasers and the led. Fiber-coupled diode
lasers are focused inside a fluidics chamber to form optical traps using underfilling beams in high NA objectives.  All the light leaving from the trap is
collected by a second objective and sent to a Position Sensitive Detector which integrates the light momentum flux, measuring changes in light momentum \cite{MEnzy.smith.2003}.
The laser beams share part of their optical paths and are separated by polarization. Part of the laser light ($\simeq$ 5\%) is deviated by a pellicle 
before focusing and used to monitor the trap position (Light Lever). Each trap is moved by pushing the tip of the optical fiber by piezos coupled to a brass tube  (wigglers).

\subsection*{Figure 2}

{\bf Coupled fluctuations in a plane.} A) A particle (P) is constrained by two springs which are oriented along the $y,z$ coordinate axes (shown by the two arrows). $k_y,k_z$ denotes the 
spring stiffnesses along $y,z$. B) A particle (P) constrained by two springs that are misaligned by an angle $\theta$ with respect to the coordinate axes. 
C) Level curves and marginal distributions for the joint equilibrium probability distribution of the particle position in the two systems shown in panels A (solid curves) and B (dashed curves). If the two springs
have very different stiffnesses, the level curves form highly eccentric ellipses. In this situation a rotation by a small angle $\theta$  in the joint distribution can lead to a large change
in the marginal distribution for $y$. The large difference in stiffness acts as a lever arm amplifying the strength of the coupling.

\subsection*{Figure 3}
{\bf Misaligned experimental configurations.} A) The coordinate system used throughout the text. The direction of light propagation (black horizontal arrows) defines the optical ($\hat z$) axis. 
The stretching direction, perpendicular to $\hat z$, defines the $\hat y $ axis.
The positions of the beads ($(y_1,z_1)$ and $(y_2,z_2)$) are measured with respect to the equilibrium positions and 
$r_0$ denotes the mean separation between the centers of the beads.
B) Aligned configuration, the tether is perfectly oriented along the $\hat y$ axis. C) Misaligned tether, the two traps are focused at different positions
along the optical axis and the tether forms an angle $\theta$ with the $\hat y$ axis. D) Misaligned traps, the principal axes of the traps form an angle $\theta$
with the $y-z$ reference frame. E) The value of $\alpha$, Eq. \eqref{gg}, as a function of the mean force, for different tethers (3kbp and 24kbp ds-DNA) and trap stiffnesses.  
The continuous lines (low trap stiffness) describes a setup similar to that used for the measurements discussed in the present paper, with $k_y\simeq 0.02$ pN/nm, $k_z\simeq 0.001$ pN/nm. The dotted line (high trap stiffness) 
describes a setup ten times stiffer (such as that described in Ref. \citep{PNAS.gebhardt.2010,Nature.comstock.2011}). The shaded area denotes the values of $\alpha$ for which 
a coupling $\epsilon\simeq0.1$ causes a 10\% error ($\epsilon^2/\alpha\simeq 0.1$).

\subsection*{Figure 4}
{\bf Trap and molecular stiffness measurements.} A) A linear dumbbell model, where three elastic elements with different 
stiffnesses are arranged in series:
Trap 1 ($k_1$), Trap 2 ($k_2$) and the tether ($k_m$).
B) Measured force variance as a function of trace length. The two data sets refer to the variance in each trap, measured on a 24kbp tether and pulled at 10 pN.
Force fluctuations in each trap are the superposition of two different linear modes. 
The solid curves are fits to the expected behavior in the case of a superposition of two modes (Supp. Mat.). The good agreement between theory
and experiment shows that the effect of low frequency noise is not relevant on our experimental timescales (less than 1\%).
C) Molecular stiffness ($k_m$) measured and 
averaged over different molecules. The continuous line shows a fit to the extensible WLC model Eq. \eqref{EWLC}, 
giving a persistence length $P=52\pm 4$ nm and a stretch modulus $S=1000\pm200$ pN, consistent with what it is reported in the literature \citep{Science.smith.1996}. 
The shaded area denotes the region where misalignment is expected to be relevant. The fair points are not included in the fit and show the effect of misalignment.
D) Comparison of the stiffness values of the two traps, $k_1$ and $k_2$,
measured through Eqs. \eqref{ia},\eqref{ib},\eqref{ic} (solid symbols) with those measured by immobilizing the bead on the micropipette (open symbols), see Materials and Methods Section. 
Measurements agree within experimental errors. Note that stiffness is measured correctly even when misalignment is relevant (shaded region). This happens
beacause the measurement of trap stiffness is mostly based on the center of mass coordinate which is not affected in the case of tether misalignment.

\subsection*{Figure 5}
{\bf Time correlation functions in ds-DNA tethers of varying contour length at different forces.} Time correlation functions were measured 
along the $y$-axis ($\langle y_+(0)y_+(t)\rangle$, $\langle y_-(0)y_-(t)\rangle$) and normalized by the variances 
($\langle y_+^2\rangle$, $\langle y_+^2\rangle$). 
Upper panels: correlation function for fluctuations of the center
of mass. The correlation function shows a simple exponential decay as expected for a single component noise. The correlation function changes with force due to trap
nonlinearity (change in trap stiffness) but does not show dependence on the length of the tether. Lower panels: correlation function for the distance between the centers 
of the beads. These correlation functions show a double--exponential behavior (Fig. \ref{Fig:3}) which denotes the presence of two relaxational processes. Data for 58 bp and 1.2 kbp DNA tethers are not shown
at 16 pN as these experiments were carried out on tethers with an inserted hairpin which unfolds around 14 pN \citep{BiophysJ.forns.2011}, and the released
ss-DNA would affect the stiffness measurement.

\subsection*{Figure 6}
{\bf Fast and slow components of the correlation function of the differential coordinate.} Double exponential fits to $\langle y_-(0)y_-(t) \rangle$ in semi log plot. Dots show the experimental data, 
the continuous fair curve superimposed on the data shows a double exponential fit to
the measured data. The dark solid curves show the fast and slow components of the double exponential fit. Every plot reports the value of $\epsilon$,
the coupling strength as obtained from Eq. \eqref{rmod}. As the molecules get shorter, the relative weight of slow fluctuations increases indicating a stronger coupling.

\subsection*{Figure 7}
{\bf Measurement of the molecular stiffness.} Main Figure: measured molecular stiffness for four tethers 58bp (triangles), 1.2 kbp (diamonds), 3 kbp (squared), 24 kbp (circles), as a function of the mean
force along the tether. Symbols are measured quantities, data from at least three different molecules have been averaged in the four cases. 
Solid lines are an EWLC fit to the data (main text). The Marko-Siggia approximation is not valid in the case of the shortest (58 bp) molecule. In this case the fit is only meant
to compare the results obtained in the DTOT with those reported in Ref. \citep{BiophysJ.forns.2011} obtained in a STOT. 
The fit results are shown in Table \ref{tab.1}. Upper right panel, comparison of the measured persistence length (P) of the three longer tethers to the empirical scaling law
proposed by Seol {\it et al.} Eq. \eqref{pheno} (solid line). Fit parameters are discussed in the Main Text. 
Lower right panel: the measured stretch modulus for the three longer tethers. Errors in $P$ and $S$ values are standard deviation
over at least 3 different molecules. 

\clearpage

\clearpage

\clearpage

\clearpage

\clearpage

\clearpage

\clearpage

\clearpage
\section*{Supplementary Information}
\setcounter{section}{0}

\section{Derivation of equations (36-41) in the Main text}

If misalignment is negligible the experimental setup is described by two coordinates $y_1,y_2$, as discussed in the Main Text. 
The equilibrium probability distribution for $y_1,y_2$ is a Gaussian distribution:
\begin{equation}
P(y_1,y_2)=\frac{1}{Z}\exp \left( \frac{(y_1,y_2)\cdot \tnsr{K}' (y_1,y_2)}{k_BT}\right), 
\end{equation}
so that the covariance matrix is easily obtained as:
\begin{equation}\label{ciolo}
\begin{split}
 \left( \begin{array}{ c c } \langle y_1^2 \rangle & \langle y_1 y_2\rangle \\
                    \langle y_1 y_2\rangle &  \langle y_2^2 \rangle
                   \end{array} \right)&=k_BT\tnsr{K}'^{-1}\\
                   &=\frac{k_BT}{k_1k_2+k_m k_1+k_2 k_m} \left( \begin{array}{ c c } k_2+k_m  & k_m \\
                   k_m &  k_1+k_m \end{array} \right).
\end{split}
\end{equation}
If we set:
\begin{equation}
\kappa=k_1k_2+k_m k_1+k_2 k_m             
\end{equation}
then, from \eqref{ciolo} we get:
\begin{eqnarray}
 \frac{k_1}{\kappa} &=&\frac{\langle y_2^2\rangle-\langle y_1 y_2\rangle}{k_BT},\label{p1s}\\
\frac{k_2}{\kappa} &=&\frac{\langle y_1^2\rangle-\langle y_1 y_2\rangle}{k_BT},\label{p2s}\\
\frac{k_m}{\kappa} &=&\frac{\langle y_1 y_2\rangle}{k_BT}\label{p3s}.
\end{eqnarray}
Moreover, using the identity:
\begin{equation}
 \frac{1}{\alpha}=\frac{\alpha}{\alpha^2}=\frac{k_1k_2}{\alpha^2}+\frac{k_1k_m}{\alpha^2}+\frac{k_mk_2}{\alpha^2}
\end{equation}
we get:
\begin{equation}
 \kappa^{-1}=\frac{\langle y_1^2\rangle \langle y_2^2\rangle- \langle y_1 y_2\rangle ^2}{(k_BT)^2}.
\end{equation}
In experimental set-ups where forces are measured directly, it is convenient to extract the trap stiffness on the basis of force fluctuation
measurements. Force and bead positions have an affine relation:
\begin{equation}\label{ggtg}
f_i=k_i y_i+f^0_i, 
\end{equation}
where $f_i^0$ is the mean tension measured in trap $i$. This affine relation can be put in a matrix form as:
\begin{equation}\label{cffc}
 \vct{f}=\tnsr{k}_D \vct{y}+\vct{f}_0
\end{equation}
with $\vct{f}=(f_1,f_2)$, $\vct{y}=(y_1,y_2)$ and
\begin{equation}
\tnsr{k}_D=\left(
 \begin{array}{cc} k_1 & 0\\ 0 & k_2
\end{array}
\right).
\end{equation}
Given the affine relations, Eq. \eqref{ggtg},\eqref{cffc}, the force covariance matrix for the force is now given by:
\begin{equation}\label{gotg}
\begin{split}
\left(\begin{array}{cc} \sigma_{11}^2 & \sigma^2_{12} \\ \sigma_{12}^2 & \sigma_{22}^2 \end{array}\right)&=k_BT \tnsr{k}_D\left(\tnsr{K}'\right)^{-1}\tnsr{k}_D=\\
&=k_BT \left(
 \begin{array}{cc}
  \frac{(k_2+k_m)k_1^2}{k_1 k_2 + k_1
    k_m + k_2 k_m} & \frac{  k_1k_2k_m}{k_1 k_2+k_1 k_m+k_2 k_m}
    \\
  \frac{k_1k_2k_m}{k_1 k_2+k_1
    k_m+k_2 k_m} & \frac{
    (k_1+k_m)k_2^2}{k_1 k_2+k_1
    k_m+k_2 k_m}
 \end{array}
 \right),
\end{split}
\end{equation}
with $\sigma^2_{ij}=\langle f_if_j\rangle-\langle f_i\rangle \langle f_j\rangle,\,\, i=1,2$.
Using Eq. \eqref{gotg} it is easy to show that:
\begin{eqnarray}
k_1&=&\frac{\sigma_{11}^2+\sigma_{12}^2}{k_B T}\label{da}\\
k_2&=&\frac{\sigma_{22}^2+\sigma_{12}^2}{k_BT}\label{db}
\end{eqnarray}
These formulae can be used to invert any element of the covariance matrix to get $k_m$:
\begin{equation}
k_m=\frac{1}{k_BT}\frac{\sigma_{12}^2\left( \sigma_{11}^2+\sigma_{12}^2\right)\left(  \sigma_{22}^2+\sigma_{12}^2\right)} { \sigma_{11}^2\sigma_{22}^2-\sigma_{12}^4}
\end{equation}

\section{Experimental Variance as a function of measurement length}

The power spectrum of a fluctuating linear mode, $x$ (Ornstein-Uhlembeck process) is:
\begin{equation}
S(\omega)=\sigma\frac{2\omega_c}{\pi (\omega^2+\omega_c^2)},
\end{equation}
where $\sigma$ is the variance $\langle \delta x^2 \rangle$ of the process and $\omega_c$ its corner frequency.
Integrating the power spectrum in the range between the inverse of the acquisition time $T$ and
the acquisition bandwidth $B$ yields the expected variance:
\begin{equation}
\langle \delta x^2\rangle_{B,T}=\int_{2\pi/T}^{2\pi B}d\omega S(\omega)=\frac{2\sigma}{\pi}\left(\arctan\left(\frac{2\pi B}{\omega_c}\right)-\arctan\left(\frac{2\pi}{T\omega_c}\right) \right) 
\end{equation}
If the acquisition bandwidth is much larger than the corner frequency ($B\gg\omega_c$) this can be approximated as:
\begin{equation}
\langle \delta x^2\rangle_{T}=\sigma\left(1-\frac{2}{\pi}\arctan\left(\frac{2\pi}{T\omega_c}\right) \right).
\end{equation}
If a signal $y$ is the superposition of two linear modes with variances $\sigma_1,\sigma_2$ and corner frequencies $\omega_1,\omega_2$ the expected behavior
for the variance $\langle \delta x^2\rangle_{T}$ as a function of the acquired trace is:
\begin{equation}
 \langle \delta y^2\rangle_{T}=\sigma_1\left(1-\frac{2}{\pi}\arctan\left(\frac{2\pi}{T\omega_1}\right) \right)+\sigma_2\left(1-\frac{2}{\pi}\arctan\left(\frac{2\pi}{T\omega_2}\right) \right),
 \end{equation}
which is the form of the fit used in Figure 5B.

\section{Dumbbell dynamics}
The discussion in the main text shows that, in absence of misalignment, using the differential and center of mass coordinates, the stiffness tensor is diagonalized
and the four fluctuation modes are uncoupled. In non-ideal cases the decoupling is not complete, but the four dimensional problem is reduced into independent
lower dimensional problems. Both in the case of trap and of tether misalignment the dynamics of the center of mass is decoupled from that of the
differential coordinate: the off diagonal terms couple either $y_-$ with $z_-$ or $y_+$ with $z_+$ but never $y_-$ with $z_+$ or $y_+$ with $z_-$.
This fact does greatly simplify the description of the dynamics of the dumbbell in Fig. 3A of the main text:
instead of considering a four dimensional problem we can consider two independent two dimensional problems: 
\begin{eqnarray}
 \dot{\vct{R}}&=&\tnsr{\mu}_{R}\left(-\tnsr{K}_+\vct{R}+{\bf \eta}_R\right),\label{motsi}\\
\dot{\vct{r}}&=&\tnsr{\mu}_{r}\left(-\tnsr{K}_-\vct{r}+\vct{\eta}_r\right) \label{motasi}.
\end{eqnarray}
Here we arranged the coordinates in two vectors: $\vct{R}=(y_+,z_+),\vct{r}=(y_-,z_-)$, $\tnsr{K}_+$ is the subtensor of $\tnsr{K}$ which affects $y_+$ and
$z_+$, and $\tnsr{K}_-$ is the subtensor which affects $y_-$ and
$z_-$. For example, in the case of tether misalignment:
\begin{equation}\label{fortsx}
 \tnsr{K}_-= \bordermatrix{ ~    & y_-     &   z_-           \cr
                          y_-   & k_y+2u(\epsilon)      &   \epsilon w(\epsilon) \cr                          
                          z_-    & \epsilon w(\epsilon) &   k_z+2 v(\epsilon)    \cr                                
                          }.
\end{equation}
with,
\begin{eqnarray}
u(\epsilon)&=&k_m (1-\epsilon^2)+\frac{f}{r_0} \epsilon^{2}\\
v(\epsilon)&=&\frac{f}{r_0} (1-\epsilon^2)+k_m \epsilon^2\\
w(\epsilon)&=&\left(k_m-\frac{f}{r_0}\right) \sqrt{1-\epsilon^2},
\end{eqnarray}
and
\begin{equation}
 \tnsr{K}_+= \bordermatrix{ ~    & y_+     &   z_+          \cr
                          y_+   & k_y      &   0 \cr                          
                          z_+    & 0 &   k_z   \cr                                
                          }.
\end{equation}
Moreover $\tnsr{\mu}_{R},\tnsr{\mu}_r$ are tensors 
describing both viscous friction on each particle and hydrodynamic interactions while ${\bf \eta}_R,{\bf \eta}_r$ are Gaussian noises with zero mean and 
correlations:
\begin{eqnarray*}
\langle  {\bf \eta}_R(t){\bf \eta}_R(s)\rangle= 2k_BT \tnsr{\mu}_{R} \delta(t-s)\\
\langle  {\bf \eta}_r(t){\bf \eta}_r(s)\rangle= 2k_BT \tnsr{\mu}_{r} \delta(t-s).
\end{eqnarray*}
After Bachelor \cite{JFluidMech.batchelor.1976} we set:
\begin{equation}
\begin{split}
 \tnsr{\mu}_{R}&=\left(\gamma^{-1}+\Gamma^{-1}\right) \frac{\vct{r}_0\otimes \vct{r}_0}{r_0^2}+\\
&+\left(\lambda^{-1}+\Lambda^{-1}\right)\left(I-\frac{\vct{r}_0\otimes \vct{r}_0}{r_0^2} \right)
\end{split}
\end{equation}
and
\begin{equation}
\begin{split}
 \tnsr{\mu}_{r}&=\left(\gamma^{-1}-\Gamma^{-1}\right) \frac{\vct{r}_0\otimes \vct{r}_0}{r_0^2}+\\
&+\left(\lambda^{-1}-\Lambda^{-1}\right)\left(I-\frac{\vct{r}_0\otimes \vct{r}_0}{r_0^2} \right),
\end{split}
\end{equation}
where $\lambda,\gamma,\Lambda,\Gamma$ are scalar parameters depending on $r_0$. In brief, $\gamma\,(\lambda)$ is the hydrodynamic friction coefficient for motions along (perpendicular to) $\vct{r}_0$,
while $\Gamma\,(\Lambda)$ is the intensity of hydrodynamic interactions along (perpendicular to) $\vct{r}_0$ (the vector connecting the centers of the beads in Fig.3A of the main text).
It is important to bear in mind that $\tnsr{\mu}_R,\tnsr{\mu}_r,k_y,k_z,k_m, \vct{r_0}$ are functions of the trap--to--trap distance $\vct{R}^T$ or, equivalently, of the mean
tension along the tether.
 The equilibrium probabilities generated by \eqref{motsi},\eqref{motasi} are given by the Boltzmann distribution i.e.: 
\begin{eqnarray}\label{distr}
 Q_{eq}(\vct{R})&=&\frac{1}{Z_R}\exp\left(-\frac{U(\vct{R})}{k_BT}\right),\\
 P_{eq}(\vct{r})&=&\frac{1}{Z_r}\exp\left(-\frac{V(\vct{r})}{k_B T}\right),
\end{eqnarray}
with
\begin{eqnarray}
U(\vct{R})&=&\frac{1}2 \vct{R}\cdot \tnsr{K}_+\vct{R},\label{u1}\\
V(\vct{r})&=&\frac{1}2 \vct{r}\cdot 2\tnsr{K}_- \vct{r}\label{u2}
\end{eqnarray}
and $Z_R,Z_r$ partition functions.
The variance of equilibrium fluctuations in $\vct{R}$ and $\vct{r}$ is connected to the elastic properties of traps and tether by:
\begin{equation}\label{pinollo}
\langle \vct{R}\otimes\vct{R}\rangle=\tnsr{K}_+^{-1} k_BT,\qquad \langle \vct{r}\otimes\vct{r}\rangle=\left(2\tnsr{K}_-\right)^{-1}k_BT.
\end{equation}
Information about hydrodynamic interactions can be obtained from the time-dependent correlation functions (tensors) of $\vct{R}$ and $\vct{r}$:
\begin{eqnarray}
\tnsr{C}_R(t)=&\langle\vct{R}(t)\otimes\vct{R}(0)\rangle\label{CR}\\
\tnsr{C}_r(t)=&\langle\vct{r}(t)\otimes\vct{r}(0)\rangle,\label{Cr}
\end{eqnarray}
which characterizes the decay of fluctuations and allows to distinguish the presence of different contributions to the total variance.
The computation of the correlation functions yields:
\begin{eqnarray}
\frac{\tnsr{C}_R(t)}{k_BT}&=&e^{-\tnsr{\mu}_R\tnsr{K}_+t}\tnsr{K}_+^{-1} \label{ctsym}\\
\frac{\tnsr{C}_r(t)}{k_BT}&=&e^{-\tnsr{\mu}_r\left(2\tnsr{K}_-\right)t}\left(2\tnsr{K}_-\right)^{-1} \label{ctasym}.
\end{eqnarray}

\section{Analysis of fluctuations: the uncoupled case $\epsilon=0$}\label{noco}

The simplest and most desirable experimental condition is that in  which the tether is perfectly aligned to the $y$ axis ($\epsilon=0$, Fig.3B). In this case fluctuations 
along the two axis are uncoupled. In the model this corresponds to the vanishing of all off-diagonal elements in the hydrodynamic and elastic tensors. 
Indeed, when $\epsilon=0$,
\begin{eqnarray}
\tnsr{\mu}_{R}&=&\left(\begin{array}{c c} \gamma^{-1}+\Gamma^{-1} & 0 \\
                       0 & \lambda^{-1}+\Lambda^{-1}
                      \end{array}\right),\\ 
\tnsr{\mu}_{r}&=&\left(\begin{array}{c c} \gamma^{-1}-\Gamma^{-1} & 0 \\
                       0 & \lambda^{-1}-\Lambda^{-1}
                      \end{array}\right),\\
\tnsr{K}_{+}&=&\left(\begin{array}{c c} k_y & 0 \\
                       0 & k_z
                      \end{array}\right)\\
\tnsr{K}_{-}&=&\left(\begin{array}{c c} k_y+2k_m & 0 \\
                       0 & k_z+2\frac{f}{r_0}
                      \end{array}\right).
\end{eqnarray}
The correlation functions are also diagonal in this case:
\begin{eqnarray}\label{saprofita}
\frac{\tnsr{C}_R(t)}{k_BT}&=&\left(\begin{array}{c c} \frac{e^{-\nu_+ t}}{k_y} & 0\\
 0 & \frac{ e^{-\nu_- t}}{k_z} \end{array}\right)\\ \label{sapro2}
\frac{\tnsr{C}_r(t)}{k_BT}&=&\left(\begin{array}{c c} \frac{e^{-\omega_+ t}}{k_y+2k_m} & 0\\
 0 & \frac{ e^{-\omega_- t}}{k_z+2f/r_0} \end{array}\right).
\end{eqnarray}
The above expressions shows the presence of 4 different frequencies in the fluctuation spectrum: 
\begin{eqnarray}
\nu_+&=&\left(\frac{1}\gamma+\frac1 \Gamma\right)k_y\\
\nu_-&=&\left(\frac{1}\lambda+\frac{1}\Lambda \right)k_z\\
\omega_+&=&\left(\frac{1}\gamma-\frac1 \Gamma\right)\left(k_y+2k_m\right)\\
\omega_-&=&\left(\frac{1}\lambda-\frac{1}\Lambda \right)\left(k_z+2\frac{f}{r_0}\right)\label{fandango}.
\end{eqnarray}

From the measurement of $\tnsr{C}_R(t)$ and $\tnsr{C}_r(t)$ it is possible to obtain the stiffness of both traps 
and molecule:
\begin{eqnarray}
 k_y&=&\frac{k_BT}{\left(\tnsr{C}_R(0)\right)_{yy}}=\frac{k_BT}{\left(\tnsr{\sigma}^2_R\right)_{yy}}\\
 k_m&=&\frac{1}{2}\left(\frac{k_BT}{\left(\tnsr{C}_r(0)\right)_{yy}}\right).
\end{eqnarray}
The time correlation function for fluctuations of $\vct{R}$ and $\vct{r}$, Eq. \eqref{saprofita},\eqref{sapro2} carries further information regarding hydrodynamic interactions, which can
be retrieved once the stiffnesses $k_y,k_m$ are known:
\begin{eqnarray}
\frac{1}{\gamma}+\frac{1}{\Gamma}&=&-\frac{1}{k_y}\frac{d}{dt} \log(\left(\tnsr{C}_R\right)_{yy})\bigg|_{t=0}\label{pun}\\
 \frac{1}{\gamma}-\frac{1}{\Gamma}&=&-\frac{1}{2k_m}\frac{d}{dt} \log(\left(\tnsr{C}_r\right)_{yy})\bigg|_{t=0}\label{mun}.
\end{eqnarray}

\section{Analysis of fluctuations with tether misalignment $\epsilon\neq 0$}

In presence of tether misalignment ($\epsilon\neq 0$) we have:
\begin{eqnarray}
 \tnsr{K}_{+}&=&\left(\begin{array}{c c} k_y & 0 \\
                       0 & k_z
                      \end{array}\right)\\
  \tnsr{K}_-&=& \bordermatrix{ ~    & y_-     &   z_-           \cr
                          y_-   & k_y+2u(\epsilon)      &   \epsilon w(\epsilon) \cr                          
                          z_-    & \epsilon w(\epsilon) &   k_z+2 v(\epsilon)    \cr                                
                          }.                     
\end{eqnarray}

\begin{equation}\label{Rmod}
\langle y_+^2\rangle=k_BT\left(\tnsr{K}_+^{-1}\right)_{y_+y_+}=\frac{k_BT}{k_y}
\end{equation}

\begin{equation}\label{rmods}
\begin{split}
\langle y_-^2\rangle&=k_BT\left(2\tnsr{K}_m^{-1}\right)_{yy} =\\
=&\frac{k_BT}{2k_m}\left(1+\left(\frac{r_0 k_m}{f}-1\right)\epsilon^2\right)+\mathcal{O}(\epsilon^3).
\end{split}
\end{equation}
Since we will be interested in tether misalignment for short tethers ($\leq 3$ kbp), in the last expression we have neglected the
trap stiffness with respect to the tether stiffness:  ($k_m\gg k_y,k_z;f/r_0\gg k_z$).
Note that whereas the variance of $\vct{R}$ is not affected by the coupling $\epsilon$, the variance of $\vct{r}$ does.
The increased $\langle y_-^2\rangle$ is due to the superposition
of two contributions, one due to fluctuations in the optical plane and the other due to fluctuations along the optical axis. In order to separate these two types of fluctuations
we need to characterize their correlation function. The $\tnsr{\mu}_r$ appearing in the correlation function Eq.\eqref{ctasym} is left invariant
under a rotation of $r_0$. This is also approximately true for $\tnsr{K}_-$ if $k_m\gg k_y,k_z;f/r_0\gg k_z$.
As a consequence the correlation function in presence of coupling can be computed as a rotation of $\tnsr{C}_r(t)$
obtained in the previous section (Eq. \eqref{sapro2}).  
If we denote by $\tnsr{C}_r(t,\epsilon)$ the correlation function in presence of a coupling of strength $\epsilon$
and by $\tnsr{R}(\epsilon)$ a rotation of an angle $\theta$ ($\epsilon=\sin \theta$) we get: 
\begin{equation}
\tnsr{R}(\epsilon)=\left(\begin{array}{c c} \sqrt{1-\epsilon^2} & \epsilon \\
                         -\epsilon & \sqrt{1-\epsilon^2}
                        \end{array}\right)
\end{equation}
and
\begin{equation}\label{rotate1}
\tnsr{C}_r(t,\epsilon)=\tnsr{R}(\epsilon)^T\tnsr{C}_r(t)\tnsr{R}(\epsilon),
\end{equation}
\begin{equation}\label{rotate2}
\frac{\left(\tnsr{C}_r(t,\epsilon)\right)_{yy}}{k_BT}=\left(1-\epsilon^{2}\right)\frac{e^{-2\omega_+ t}}{2 k_m}+\epsilon^2 \frac{r_0e^{-2\omega_- t}  }{2f}.
\end{equation}

Similar although more cumbersome formulas can be obtained in more general cases, i.e. when the trap stiffness $k_y,k_z$ are comparable or larger
than $k_m,f/r_0$ respectively.
Summarizing, in presence of misalignment along the $z$ axis we expect the correlation function of the relative distance to be
a double exponential exhibiting two widely separated timescales: a fast timescale ($\omega_+^{-1}$) due to fluctuations in the optical plane and
a slow timescale ($\omega_-^{-1}$) due to fluctuations along the optical axis. 
Once the two components have been separated through fitting, as shown in the Main Text, the same analysis as in the uncoupled case can be performed on the fast component
of the correlation function, to measure the molecular stiffness and the hydrodynamic parameters.
From the slow contribution of the correlation function (second term in the r.h.s. of Eq. \eqref{rotate2}) it is also possible to extract the coupling parameter, 
since the ratio $f/r_0$ is independently known.
In all the experiments presented in this paper, the coupling parameter was not higher than 0.25, which corresponds to an angle $\theta\simeq 15^o$. 
In our setup, especially for short tethers, we can have $\epsilon^2\simeq \alpha$, making the slow contribution to the variance \eqref{rmod} comparable or even 
bigger than the one due to fast fluctuations.

\end{document}